## BUILDING DESIGN IN TROPICAL CLIMATES. ELABORATION OF THE ECODOM STANDARD IN THE FRENCH TROPICAL ISLANDS

**François Garde, Harry Boyer**
Laboratoire de Génie Industriel, University of Reunion Island, IUT de Saint Pierre, avenue de Soweto, Saint-Pierre, Reunion Island, 97 410 , France, (+262) 96 28 91, (+262) 96 28 99, garde@univ-reunion.fr

**Robert Celaire**
Concept Energie, 1 rue Mirabeau, Lambesc,13410, France, (+33) 04 42 92 84 19, (+33) 04 42 92 71 36, Robert.Celaire@wanadoo.fr

**Laurent Seauve**
Sunstice, Centre d'Affaires Actualis, Guadeloupe, (+590) 84 04 28, (+590) 84 04 28, sunstice@outremer.com

**Abstract** – This paper deals with the elaboration of global quality standards for natural and low energy cooling in french tropical island buildings. Electric load profiles of tropical islands in developed countries are characterised by morning, midday and evening peaks arising from all year round high power demand in the commercial and residential sectors, mostly due to air conditioning appliances and bad thermal conception of the building. In early 1995, a DSM pilot initiative has been launched in the french islands of Guadeloupe and Reunion through a partnership between the French Public Utility EDF, institutions involved in energy conservation, environment preservation (ADEME) and construction quality improvment, the University of Reunion Island and several other public and private partners (low cost housing institutions, architects, energy consultant, etc...) to set up a standard in the thermal conception of buildings in tropical climates. This has led to definition of optimized bioclimatic urban planning and architectural design, the use of passive cooling architectural components, natural ventilation and energy efficient systems. The impact of each technical solution on the thermal comfort within the building was evaluated with an airflow and thermal building simulation software (CODYRUN). These technical solutions have been edited in a pedagogical reference document and have been implemented in 300 new pilot dwelling projects through the year 1996 in Reunion Island and in Guadeloupe island. An experimental follow up is still in process in the first ECODOM dwellings for an experimental validation of the impact of the passive cooling solutions on the comfort of the occupants and to modify them if necessary.

## 1. INTRODUCTION

Each year 20,000 dwellings are built in the French Overseas Departments. There are four French Overseas Departments (DOM) : two islands are located in the Caribbean (Martinique and Guadeloupe), one situated 400 km to the east of Madagascar in the Indian Ocean (Reunion Island) and the fourth Department is in the North of Brazil (French Guiana). Each experiences a hot climate, tropical and humid in the islands of Guadeloupe, Martinique and Reunion and equatorial in French Guiana. Three quarters of this development is welfare housing. Initially this new housing was built without the comfort of air conditioning or hot water which has led to the haphazard installation of instant electrical hot water boilers and badly situated, thought out and maintenanced individual air conditioning systems. The lack of thermal regulations, combined with a building design widely inspired from temperated climates and the economic constraints of a tight budget for construction have led to the development of buildings totally unadapted to the tropical climate. The large population increase in the DOM, the rise in living standards, and the decreasing costs of air conditioning appliances constitute a real energetical, economical and environmental problem.

When considering the economical aspects, the high production cost also generates a continual high deficit for the French Electricity Board (EDF). EDF losses 350 millions Euros every year in overseas departments. The average selling price of electricity is less than 9 Euro cents (less then hlaf than the real production-distribution cost) because of the french pricing policy (selling price of electricity is the same as in mainland France).

All these factors point out that passive solar cooling in thermal conception of buildings is therefore of great economical, social and environmental importance. A long term overall programme to improve comfort and energy performance in residential and commercial buildings has been set up in the overseas territories. In the new housing sector, a quality standard seal has been launched concerning the building structure, the hot water production systems and the air conditioning appliances.

## 2. THE ECODOM STANDARD

A Demand Side Management pilot initiative called ECODOM was launched in early 1995 in the French islands of Guadeloupe and Reunion through a partnership between the French electricity board (EDF), institutions involved in energy saving and environmental conservation (ADEME) and construction quality improvment, the ministries of Housing, Industry and the French Overseas Departement, the University of Reunion island and several other public and private partners, such as low cost housing institutions, architects, energy consultants, etc...

*2.1 The objectives*

The ECODOM standard aims to simplify the creation of naturally ventilated comfortable dwellings whilst avoiding the usual necessity of a powered air cooling system consuming electricity. The ECODOM standard provides simple technical solutions, at an affordable price.

The existing bibliography in the passive thermal design of buildings is extremely rich and various. Publications often focus either on the optimization of one composant of the building (Bansal,1992), (Malama 1996), (Rousseau 1996) (Peuportier, 1995) or on the presentation of a performant bioclimatic project (Filippin et al, 1998), (Ashley and Reynolds, 1993) or on a global approach of the building with a description of the passive solar options to apply on it (Garg, 1991), (Hassid, 1985), (Millet, 1988), (Gandemer 1992). These publications often present an obvious interest for the building physicians but can not be easily integrated in a national overall program to improve the thermal performances of building because the scientific language is often far from the economical building reality. Architects and engineers lack of time to appropriate themselves the scientific tools and to read all the research reports in the fields of thermal design of building. Thus, to bridge the gap between building designers and building physicians, a simple, common, and pedagogical language must be spoken.

One of the objectives of the ECODOM project was to define simply rules which can be easily understood by the whole building community. A reference document comprising a pedagogical presentation of the technical passive options was published in 1996. This document allow all the building designers to speak the same language in the study of futures ECODOM buildings.

Once the document made, the next step was obviously to implement the standard to 300 pilot new dwelling projects throughout the year 1996, then, after a technical and social validation period, to expand this pilot phase in the residential sector on a much broader scale (2000 new dwellings per year), and complete similar global energy efficiency projects in existing housing and large and medium size commercial buildings.

Finally, ECODOM has also social objectives because more than half of the 20 000 dwellings in the french overseas departments are constructed by public housing companies. This allow to give minimum comfort for people who will never be able to afford air conditioning investments and subsequent electricity bills.

*2.2 The prescriptions*

Comfort level is reached through an architectural building design adapted to the local climate : the dwelling is protected from the negative climatic parameters (the sun) and favours the positive climatic factors (the wind).

The achievement of a good level of thermal comfort necessitates the application of a certain number of compulsary rules. These rules concern immediate surroundings of the dwelling and its constituent componants. They cover five points:

1) Location on site (vegetation around the building);
2) Solar protection (roof, walls, windows);

3) Natural ventilation (exploitation of trade winds, and optimized ratio of inside/outside air-permeability of the dwelling) or mechanical ventilation (air fans);
4) Domestic hot water production (servo-controlled night electric drum, sized according to requirements, solar or gas water heaters);
5) Option, air conditioned bedrooms (closed room and efficient, regulated appliances).

## 3. METHODOLOGY

To reach these quality standards, an important number of simulations were computed on each component of the building in order to quantify the thermal and energetic impact of each technical solution on the thermal comfort within the building. Various authors have already worked on specified problems concerning the outside structure of the building : (Bansal, 92) on the effect of external colour, (Malama, 96) on passive cooling strategies for roof and walls, (Rousseau, 96) on the effect of natural ventilation, (De Walls 93) for global considerations on the building adapted for a defined climate.

Our approach consisted in the study of typical dwellings, selected as the most representative type of accomodation built in the Reunion island, in terms of architecture and building material (see figure 1).

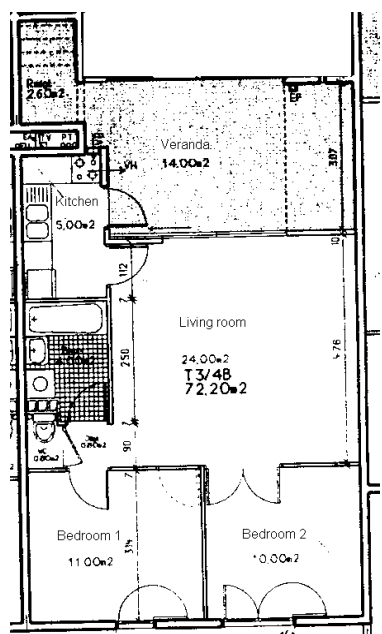

Fig. 1 : Typical dwelling

The simulations were carried out with the use of a building thermal and airflow simulation software on the constituant components (roof, walls, windows) and on natural ventilation, in a way to estimate the influence of each of the above prescriptions, in terms of thermal comfort and energetic performances. This has led to the definition of performant passive technical solutions for each comprising part of the structure and likewise a minimum ratio to optimise the natural ventilation. These simulations, their analysis and the synthesis of the results have been presented in (Garde, 99).

## 4. ECODOM : THE REFERENCE DOCUMENT

We will illustrate in the following the final results which are presented in the ECODOM reference document (Célaire, 1997). This reference document is very helpful and essential because it is used as a tool for the study of the new ECODOM projects between the architects, the building physicians and the engineers.

### 4.1 Position on site

Performant thermal and energetic housing conception starts immediately from their location on the building site. The immediate surroundings of a building have a significant influence on the conditions of thermal comfort inside. This is particularly the case of the surrounding surface of the building, which should neither reflect the solar radiations towards the house nor increase the ambient air temperature.

The results concerning the surroundings are :
The finished surface around the building should be protected from direct sunlight on more than three quarters of its perimeter, with a width of at least 3 metres. This can be satisfied by either vegetation (lawn, bushes, flowers) around the building, or by all vegetation sun-blocks. These prescriptions are similar to the recommendations of (De Wall, 1993) concerning urban planning for warm humid climates.

### 4.2 Solar protection

In a humid tropical climate, the sources of uncomfort arise from a temperature increase due to bad architectural conception, concerning insulation. 80% of this is due to solar radiation, the rest, to conduction exchanges. The setting up of an efficient solar protection constitutes the second fundamental phase in the thermal design of buildings. This protection concerns all the exterior separations of the dwelling : roof, walls and windows.

### Solar protection of the roof

Thermal inflows represents up to 60% of the overall inflows from the separations in the dwellings. An efficient solar protection for the roof is therefore of prime urgence for a good thermal conception.

The following table is valid for terrasse rooves, inclined roofs without lofts, rooves with closed or barely ventilated lofts.

Well ventilated lofts should have ventilation ducts spread out uniformely throughout the perimeter, which surface conforms to the following equation :

$$\frac{S_0}{S_t} = \frac{Total\ area\ of\ openings}{Roof\ area} \geq 0.15 \qquad (1)$$

In this case, the ceiling under the loft should satisfy certain prescriptions (see table 1).

Table 1 only gives results for polystyrène and polyurethane, as these kinds of insulation are the most commonly used at an affordable price under humid tropical climates. Other alternatives exist and can be used if their equivalent thermal resistance is sufficient compared to the results given in Table 1.

Table 1
Roof solar protection

| Insulated simple roofs | | |
|---|---|---|
| Roof colour | Polystyrène type insulation $\lambda = 0.041$ W/m.K | Polyuréthane type insulation $\lambda = 0.029$ W/m.K |
| light ($\alpha = 0.4$) | 5 cm | 4 cm |
| medium ($\alpha = 0.6$) | 8 cm | 6 cm |
| dark ($\alpha = 0.8$) | 10 cm | 8 cm |
| Roofs with well-ventilated lofts | | |
| Roof colour | Polystyrène type insulation $\lambda = 0.041$ W/m.K | Polyuréthane type insulation $\lambda = 0.029$ W/m.K |
| dark ($\alpha = 0.4$) | No insulation needed | |
| medium ($\alpha = 0.6$) | 2 cm | 0 cm |

### Solar protection of walls

The thermal gains from the walls represents 20 to 30% (40 à 65 % for the dwellings which are not under the roof) of the thermal gains from the separations. Various solutions enable a protection of the walls from the sunlight : horizontal or vertical canopy or overhang, thermal insulation of the walls. The results obtained from the simulations constitute the following table, which give the optimum dimensions of the canopy in relation to the orientation of the walls and the walls inertia for Reunion island.

Table 2
Overhang of canopy - minimum d/h ratio values to be respected.

| Type of wall | light colour | | | | medium colour | | | |
|---|---|---|---|---|---|---|---|---|
| | East | South | West | North | East | South | West | North |
| concrete 20cm (R=0.1 m².K/W) | 0.4 | 0.2 | 0.7 | 0.5 | 1 | 0.5 | 1.3 | 0.7 |
| hollow concrete blocks (R=0.2 m².K/W) | 0.1 | 0.1 | 0.3 | 0.2 | 0.5 | 0.3 | 0.8 | 0.5 |
| wood (R=0.5 m².K/W) | 0 | 0 | 0 | 0 | 0 | 0 | 0.2 | 0.1 |

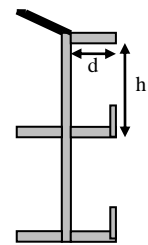

Table 3
Insulation of walls (in cm) for different orientations and external colours (for a conductivity of 0.041 W/m.K )

| Type of wall | light colour | | | | medium colour | | | |
|---|---|---|---|---|---|---|---|---|
| | East | South | west | North | East | South | West | North |
| concrete 20cm (R=0.1 m².K/W) | 1 | 1 | 1 | 1 | 2 | 1 | 2 | 2 |
| hollow concrete blocks (R=0.2 m².K/W) | 1 | 1 | 1 | 1 | 1 | 1 | 2 | 2 |
| wood (R=0.5 m².K/W) | 0 | 0 | 0 | 0 | 0 | 0 | 1 | 1 |

When the walls do not have a canopy, the minimum thicknesses of insulation (in cm) needed for the different types of walls and in different orientations are shown in Table 3.

If the values of d/h seem excessive, other alternative such as vertical shading with an airgap may be considered. In that case, no recommandation are needed for each orientation of the wall. Meanwhile, these values are fully compatible with the creole architecture components such as veranda and balcony which insure an efficient solar protection of walls and windows.

*Solar protection of windows :*

The protection of the windows is fundamental, not only because they represent 15 to 30% of the thermal gains but also because they contribute to the increase in the uncomfort experienced by the occupant, due to the instant heating of the ambiant air temperature and an exposion to direct or reflected sunlight. All the windows must therefore be protected by some sort of window shading, such as horizontal canopies and other shading devices such as venitian blinds or opaque, mobile strips. The simulations enabled us to optimize the geometric characteristics of the horizontal canopies in relation to the orientation of the glazing. (see table 7).

Table 4
Values of d/(2a+h) (case 1), or d/h (case 2)

| | Orientation of windows | | | |
|---|---|---|---|---|
| | East | South | West | North |
| Reunion Island | 0.8 | 0.3 | 1 | 0.6 |

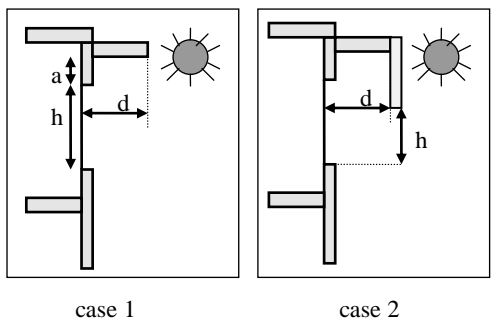

case 1    case 2

*4.3 Natural ventilation*

$$P1 = \frac{So1}{Sp} \geq 0.25$$

$$P2 = \frac{So}{Sp} \geq 0.25$$

$$Sp = \frac{Sp1 + Sp2}{2}$$

$$Si1 \geq So1 \text{ or } So2$$

$$Si2 \geq So1 \text{ or } So2$$

In warm climates, natural ventilation is the most usual means of heat transfer from both occupants and buildings : The natural ventilation, depending on its importance, ensures three functions :

☐ Weak flow (1 to 2 vol/h) for the preservation of hygiene conditions by air renewal;

☐ Moderate flow (40 vol/h), for the evacuation of internal gains and the cooling of the outside structure;

☐ High flow (more than 100 vol/h) to assure the comfort by sudation.

Thus the high air speed and its good layout betters the sudation process. This is the only means which enables the compensation of the high temperatures, coupled to a high rate of hygrometry.

Our aim is therefore to find the exterior/interior permeability coupling which enables us to obtain the rate of air renewal of 40 vol/h. On the one hand the structure of the dwelling will be sufficiently cooled and on the other, such an air renewal rate allows us to hope for wind speeds of 0.2 à 0.5 m.s-1 , which is largely sufficient, when taking into account the climatic parameters (outside temperature rarely greater than 32°C), to assure a good level of comfort.

We found from the simulations that the critical air renewal rate of 40 vol/h is obtained for a configuration of exterior permeability equal to 25% and interior permeability of 25%, equally for a light structure, as for a heavy structure. The natural ventilation is simply more effective during the night for the heavy structure, whereas in the light structure it serves mainly to evacuate the overheating from during the day.

Thus the dwelling should have complete cross ventilation (see Fig. 2). At each level or floor, there should exist openings in the principal rooms, on at least two opposing facades (the principal rooms being the bedroom and the living room). Also the interior lay-out should be designed in a way that the outside air, flows through the principal rooms and the corridoors, from one facade to the other, by the doors and the other openings in the partitions.

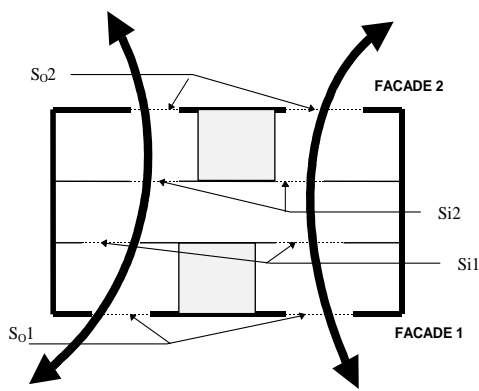

Fig. 2 : Cross ventilated dwelling

The details of calculation needed to determine the exterior and interior permeabilities of 25% are exposed below. The first thing to do is to calculate the mean of the permeabilities of the two opposite facades of the building. The ECODOM permeability (internal and external) required is 25% of the mean.

So1 : Net surface of external openings, principal rooms (façade 1);
So2 : Net surface of external openings, principal rooms (façade 2);
Sp1, Sp2 : Total surface of principal rooms of façades 1 and 2;
Si1 and Si2 : Total surface of internal openings.

**Air fans :**

When natural ventilation air speed is unsufficient, air fans can additionally be used. This enables an increase in the comfort range of more than 2°C (Fuad, 1996).

Each room in the dwelling should be equipped with electric wiring in the ceiling, wired to a wall switch, destined exclusively for the installation of air fans.

**4.4 Air conditioned bedroom option:**

In certain dwellings, and at certain times of the year, natural ventilation, even with the existence of air fans, is not adequate for an acceptable level of comfort. In this case we can choose to air condition the bedrooms using efficient appliances. The simulations which we carried out show that the air conditioning charges can be reduced through good structure conception and control of the air renewal rate. For a light structure these savings reach 3.4 cooling kWh per day, and 11 kWh for a heavy structure, where the inertia plays a dominating role in the air conditioners consumption. Throughout the whole of the wet season, the consumption was diminished by half with a good structure conception (1000 cooling kWh). The maximum cooling power is therefore 80W/m².

Practically, the air conditioners should meet certain standards of efficiency (cooling efficiency of 2.5 for the window units and 3 for split-systems), of permeability (each room should be equipped with a mechanically controlled air renewal of 25m3/h) and a maintenance contract.

**4.5 Domestic hot water**

We have not carried out simulations in this domain, however a certain number of prescriptions need to be verified, and we feel that they need to be specified as water heating consumes greatly and constitutes a real energy problem. It is important that the dwellings are equipped with efficient long-lasting and economic, domestic hot water heating systems. The water heater can be solar, electric or gas.

In the case of solar water heaters, the apparatus must conform to the technical control CSTB. The total minimum surface of the solar captors, should be installed in relation to the size of the dwelling.(see table 8). The capacity of the water storage should be 60 and 120 litres per square metre per net square metre of captor. The conventional minimum annual production should be 700kWh per net square metre of the capting surface.

Table 8
Technical characteristics - solar water heaters

| Solar water heaters minimum captor surface installed for each dwelling | |
| --- | --- |
| F1-F2 (2 rooms) | 1.5 m² |
| F3 (3 rooms) | 2.0 m² |
| F4 (4 rooms) | 2.5 m² |
| F5 (5 rooms) | 3.0 m² |
| F6 and more | 3.5 m² |

When considering electrical heaters the appliance must have the mark of the approved French standard of manufacture NF (Norme Française). The instant hot water heaters are high energy consuming, so therefore are excluded.The capacity of the water heater and its.cooling constant  are calculated in function to the number of principal rooms within the dwelling (see table 9). The power supply should be equipped with a three position commutation swich: servo-controlled to the off peak hours, over-ride, off.

The gas water heaters must have the French Standard Mark and also provision must be made for a burnt gas outlet.

Table 5
Technical caracteristics of electric water heater

| Individuel electric water heater | | |
| --- | --- | --- |
| Type of dwelling | minimum storage capacity | Cooling constant |
| F1-F2 | 100 l. | 0.32 |
| F3 | 150 l. | 0.23 |
| F4 | 200 l. | 0.22 |
| F5 | 250 l. | 0.22 |
| F6 and more | 300 l. | 0.22 |

## 5. PRESENTATION OF ONE ECODOM PROJECT : "LA DECOUVERTE"

At the present time, three experimental operations have been built in Reunion Island, and two in Guadeloupe, that is to say a total number of 300 dwellings.

One of these called "La Découverte" is represented by figure2.
44 dwellings were studied with the ECODOM prescriptions as part of the project "La Découverte"

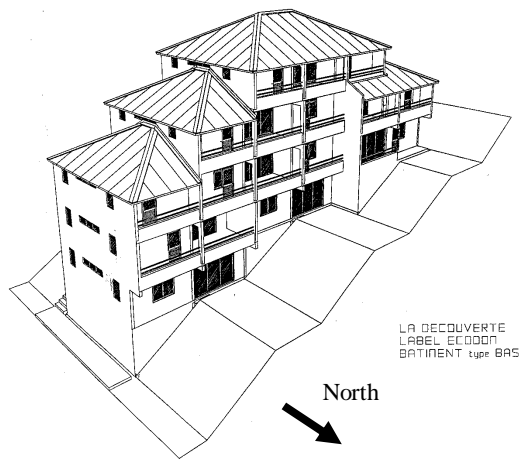

North

Fig. 3 : ECODOM "La Découverte", 44 dwellings, 1997.

This project started in 1996 when the first contacts were taken between the architect and the building owner.
The main problem was to create a confident relationship because ECODOM must be presented as a partnership with no interference with the creativity of the architect.

Figure 3 represents the project after the application of the ECODOM prescriptions
The ECODOM modifications applied on the project are explained in the following.

No modification was made for the position on site because gardens around the buildings have already been planned (see figure 4). Otherwise, the project has an attractive orientation as the main façades have a north and south orientation, with small west and east sides. Therefore, the only façade to be processed is the north one as the south one is seldom sunny (in the southern hemisphere, the sun goes to the south only in december).
Relating to the solar protection of the roof, the architect has provided 5 cm of insulation for red and blue roofs. According to table 1, it is not enough. Therefore, the architect was asked to increase the insulation thickness up to 8 cm (medium colours) and to rather use polysytyrene as type of insulation than mineral whool. This kind of insulation is very cheap but not very adapted to tropical climates : the mineral whool loses its thermal properties because of the condensation.
Regarding to the solar protection of walls, the colour is light and the materials used are hollow concrete blocks. In table 2, it is recommended to put 1 cm to the East and West sides. The south façade was not treated because of the important number of overhangs and entrances to the dwellings. Thus, as the initial project didn't provide solar protection of the bedrooms windows, is was asked to add overhangs above the exposed windows on the north facade.

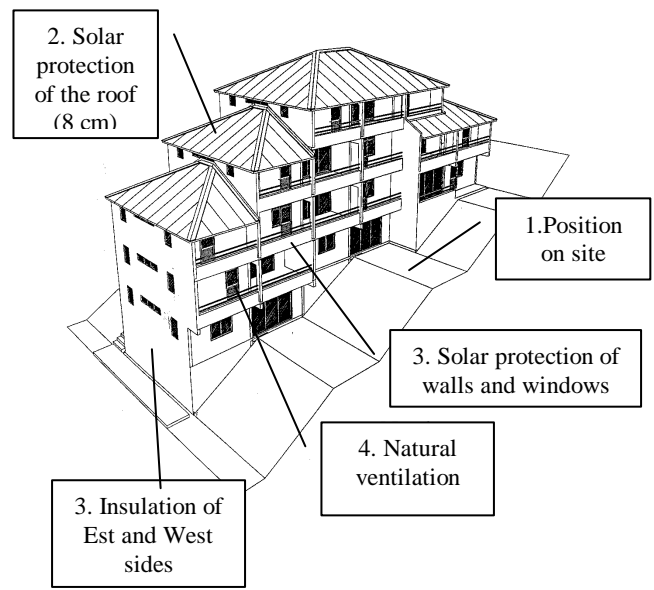

2. Solar protection of the roof (8 cm)

1.Position on site

3. Solar protection of walls and windows

4. Natural ventilation

3. Insulation of Est and West sides

Fig. 4 : Modifications of the original project

..However, the solar protection of windows is strongly correlated to the natural ventilation. The bigger is the opening, the longer must be the overhang size.

Concerning natural ventilation, the dwelling permeability in the initial project was not sufficient. ECODOM needs a permeability of 2 square meters and the initial project has only 1.44 square meter. That is why all the windows have been suppressed and replaced by door windows to increase the permeability. Balconies were also added all along the facade for the solar protection of the openings and of the walls.

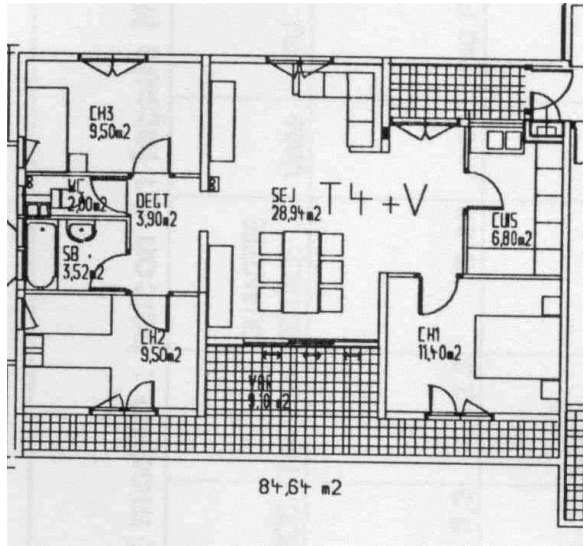

Fig. 5 : Plan of one ECODOM dwelling. Increase of the permeability of the bedrooms (door windows) and solar protection of the north facade (balcony).

## 6. DESCRIPTION OF THE SOCIAL AND TECHNICAL FOLLOW UP

An experimental follow up has been launched for the first ECODOM dwellings in order to validate experimentally the impact of the passive cooling solutions on the comfort of the occupants.

The operation was delivered in april 1999. With the agreement of the housing company, four dwellings have been instrumented during the warm season (from january to april). A portable dattalogger was set up close to the buildings to record all the climatic variables (external temperature and humidity, solar radiation, wind speed and wind direction).

This experimental period was done to estimate the impact of the prescriptions on the building thermal performances without taking into account the influence of the occupants.

A second instrumentation period is scheduled during the next warm season with the occupants inside. This second experimental step will be coupled with a social study while the people are living inside in order to understand how they feel in terms of thermal comfort, acoustic comfort (because in tropical climates, it is difficult to reach in the same time thermal comfort and acoustic comfort), visual comfort, environment.

The dwellings were selected in order to compare various parameters such as the roof and walls thermal insulation, the influence of natural ventilation.

During the non-occupied period, the probes used are thermocouples for the measurement of air temperature, resultant temperature, surface temperatures on walls and roofs, thermohygrometers for the relative humidity, accurate anemometer for the air speed inside the dwellings (see fig.6).

During the occupied period, the probes are small white boxes with internal memory which can be fixed on a wall. They give the temperature/hygrometry couple measurements each 30 minutes during several months.

By the time this article is been writting, the measurements are still in process.

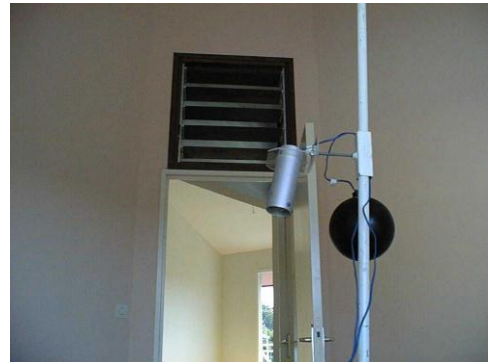

Fig. 6 : Measurements of air temperature and resultant temperature inside the dwelling

The experimental results will allow to have some feed back informations about the ECODOM dwellings and to supply the appropriate corrections on the numerical data of the passive solar solutions. Above all, this will allow to know if the dwellings are fitted to the occupant way of life

## 7. CONCLUSION

All the methodology used for the elaboration of the ECODOM standard, from the simulations to the experimental results, has been presented.

The whole work has led to define performant solar passive cooling solutions for each comprising part of the structure and likewise a minimum ratio to optimise the natural ventilation. The dwellings to be constructed following the ECODOM specifications should satisfy the criteria of these technical solutions. The simplicity of the specifications allow all the ECODOM actors (architects, building physicians, building designers) to speak the same language.

An experimental follow up has been made for the first ECODOM dwellings in order to validate experimentally the impact of the passive cooling solutions on the comfort of the occupants. This experimental period is still in process and the feed back will be available in the beginning of the year 2000.

This follow up is important, as the setting up of the ECODOM standard will be the first step towards the setting up of thermal regulations in the French overseas departments, by the year 2002.

## REFERENCES

Bansal N.K, Garg S.N., and Kothari S. (1992). Effect of Exterior Surface Color on the Thermal Performance of Buildings. *Building and Environment.* 27, 31-37.

Malama A. and Sharples S. (1996). Thermal and Economic Implications of Passive Cooling Strategies in Low-Cost Housing in Tropical Upland Climates. *Architectural Science Review.* 39, 95-105.

Rousseau P.G. and Mathews E.H. (1996). A new integrated design tool for naturally ventilated buildings. *Energy and buildings.* 23, 231-236.

Peuportier B. and Michel J. (1995). Comparative analysis of active and passive solar heating systems with transparent insulation. *Solar Energy*. 54, 1, 13-18.

Filippin C., Beascochea A., Esteves A., De Rosa C., Cortegoso L. and Estelrich D. (1998). A passive solar building for ecological research in Argentina : The first two years experience. *Solar Energy.* 63, 2, 105-115.

Ashley R. and Reynolds J.S. (1993). Overall and zonal efficiency end use in an energy conscious office building. *Solar energy*. 52, 1, 75-83

Garg N.K. (1991). Passive solar options for thermal comfort in building enveloppes – An assessment. *Solar energy*. 47, 6, 437-441.

Hassif S. (1986). A linear model for passive solar evaluation. *Solar Energy*. 37, 2, 165-174.

Millet J.R. (1988). *Conception thermique des bâtiments en climatisation naturelle dans les D.O.M. Document d'application. C.S.T.B n° 88.4343,* 124 p.

Gandemer J., Barnaud G., Sacre C., Millet J.R. (1992). *Guide sur la climatisation naturelle en climat tropical humide. Tome 1 : Méthodologie de prise en compte des paramètres climatiques dans l'habitat et conseils pratiques*. C.S.T.B., 1992. 133 p.

De Waal H.B. (1993). New Recommendations for Building in Tropical Climates. *Building and Environment*. 28, 271-285.

Garde F., Boyer H., Gatina J.C. (1999) Elaboration of global quality standards for natural and low energy cooling in french tropical island buildings. *Building and Environment*. 34, 71-83.

Célaire R. (1997). *Opération expérimentale ECODOM, cahier de prescriptions. Document de référence.*